\documentclass[twocolumn,floatfix]{revtex4}
\usepackage{graphics}

\begin{document}

\title{Correlations in photon-numbers and integrated intensities
in parametric processes involving three optical fields}

\author{Jan Pe\v{r}ina$^{1,2}$\thanks{E-mail: perina@prfnw.upol.cz},
Jarom\'\i r K\v{r}epelka$^{2}$, Jan Pe\v{r}ina Jr.$^{2}$, Maria
Bondani$^{3}$, Alessia Allevi$^{4}$, Alessandra Andreoni$^{5}$}

\affiliation{$^{1}$ Department of Optics, Palack\'{y} University,
17. listopadu 50, 772 07 Olomouc, Czech Republic \\
$^{2}$Joint Laboratory of Optics, Palack\'{y} University and
Institute of Physics of Academy of Sciences of the Czech Republic,
17. listopadu 50, 772 07 Olomouc, Czech Republic \\
$^{3}$ National Laboratory for Ultrafast and Ultraintense Optical
Science C.N.R.-I.N.F.M., Via Valleggio 11, 22100 Como, Italy \\
$^{4}$ C.N.I.S.M., U.d.R. Como, Via Valleggio 11, 22100, Italy \\
$^{5}$ Department of Physics and Mathematics, University of
Insubria and C.N.I.S.M., U.d.R. Como, Via Valleggio 11, 22100
Como, Italy}

\begin{abstract}
Two strongly-pumped parametric interactions are simultaneously
realized in a single nonlinear crystal in order to generate three
strongly correlated optical fields. By combining together the
outputs of two of the three detectors measuring intensities of the
generated fields, we obtain the joint photocount statistics
between the single field and the sum of the other two. Moreover,
we develop a microscopic quantum theory to determine the joint
photon-number distribution and the joint quasi-distributions of
integrated intensities and prove nonclassical nature of the
three-mode state. Finally, by performing a conditional measurement
on the single field, we obtain a state endowed with a
sub-Poissonian statistics, as testified by the analysis of the
conditional Fano factor. The role of quantum detection
efficiencies in this conditional state-preparation method is
discussed in detail.
\end{abstract}

%\keywords{photon-number distribution, distribution of integrated
%intensity, parametric processes, nonclassical light, entanglement
%in photon numbers}


\maketitle

\section{Introduction}

The quantum properties of photon pairs generated in the process of
spontaneous parametric down-conversion
\cite{Walls1994,Mandel1995,Perina1994} have been investigated in
many experimental and theoretical works during the last 30 years.
Pairwise character of the fields generated in this process has
been used, e.g., for testing fundamental laws of quantum mechanics
\cite{Perina1994}, as well as for quantum teleportation
\cite{Bouwmeester1997}, quantum cryptography \cite{Lutkenhaus2000}
and in metrology applications \cite{Migdal1999}. The theory that
describes this kind of interaction has been elaborated from
several points of view for fields containing just a fraction of a
photon pair \cite{Keller1997,PerinaJr1999,PerinaJr2006} and for
fields composed of many photon pairs \cite{Nagasako2002}. Also
stimulated emission of photon pairs has been addressed
\cite{Lamas-Linares2001,DeMartini2002,Pellicia2003}.

More recently, the development of the field of quantum-information
processing \cite{Bouwmeester2000} has drawn attention to
three-field quantum correlations. In fact, interesting
correlations can be reached when strongly-pumped parametric
down-conversion and parametric amplification are combined together
through a common field. Such a system can be built in a single
nonlinear crystal oriented in such a way that phase-matching
conditions for both the interactions are fulfilled together
\cite{Allevi2006}. It has been demonstrated that the tripartite
state generated in this way is endowed with entanglement in the
number of photons. In particular as the constant of motion
admitted by the hamiltonian that describes the process suggests,
we obtain that the number of photons in one of the three generated
fields is always equal to the sum of the photons in the other two
fields \cite{Allevi2008}. These properties make the system useful
for the generation of nonclassical states by means of a suitable
conditional state preparation. In more detail, if a given number $
n_0 $ of photons in field $ a_0 $ is detected then the remaining
two fields $ a_1 $ and $ a_2 $ are ideally left in a state $
|\psi\rangle = \sum_{l=0}^{n_0} c_l |l\rangle_1 |n_0-l\rangle_2 $
entangled in photon numbers, $ | l \rangle_i $ means the Fock
state with $ l $ photons in field $ a_i $ and coefficients $ c_l $
depend on the nonlinear interaction.

Here we present the experimental realization of this scheme and in
particular we study the joint photocount statistics between one
field (single field) and the sum of the other two (compound field)
in order to find out the experimental conditions in which it is
possible to realize a deterministic source of states entangled in
the number of photons. We note that this source requires
photon-number resolving detectors
\cite{Kim1999,Miller2003,Haderka2004,Rehacek2003,Achilles2004,Fitch2003}.

Note that fields composed of photon pairs have been experimentally
investigated
\cite{Agliati2005,Haderka2005,Haderka2005a,Bondani2007,Paleari2004}
under conditions that allowed having up to several thousands of
photon pairs per pump pulse. A detailed theory based on a
multi-mode description of the generated fields has been developed
for spontaneous \cite{Perina2005} as well as for stimulated
processes \cite{Perina2006} with the aim to interpret the
experimental data. In this work, as the photocounts from two of
the three detectors are combined together, we also use this theory
in order to correctly interpret and process the measurements. In
particular, we can determine the joint single-compound-field
(JSCF) photon-number distribution and the JSCF quasi-distributions
of integrated intensities from the measured JSCF photocount
distribution. Moreover, the conditional photocount and
photon-number distributions, together with the corresponding Fano
factors, are derived in order to point out the versatility of the
scheme as a source of nonclassical states. To this aim, the
quantum detection efficiency is an important parameter.

Theoretical description of the nonlinear process is presented in
Sec.~II. Photon-number distributions and quasi-distributions of
integrated intensities as well as other properties of the measured
fields are derived in Sec.~III. Experimental results are discussed
in Sec.~IV. Sec.~V gives conclusions.

\section{Three-mode parametric process}

We consider a three-mode parametric process in which two strong
coherent classical fields pump two interlinked nonlinear
interactions: a frequency down-conversion and a parametric
amplification. The corresponding interaction Hamiltonian can be
written as follows \cite{Mishkin1969,Allevi2006}:
\begin{equation}     %1
 H_{\rm int}=\gamma_0 a_0^{\dagger}a_2^{\dagger} + \gamma_1
 a_1^{\dagger} a_2 + {\rm h.c.},
\label{1}
\end{equation}
where $a_{i}$ ($a_{i}^{\dagger}$), $ i=0,1,2 $, are the
corresponding photon annihilation (creation) operators and
$\gamma_0 $ and $ \gamma_1$ are coupling constants for frequency
down-conversion and parametric amplification, respectively. Symbol
h.c. stands for the hermitian conjugated term. Analytical solutions of
the corresponding Heisenberg equations for the annihilation and creation operators are
given in \cite{Allevi2006}. They can be used for the determination
of the normal three-mode characteristic function of the
spontaneous process:
\begin{eqnarray}       %2
 C(\beta_0,\beta_1,\beta_2) &=& \exp[-B_0|\beta_0|^2-B_1|\beta_1|^2
 -B_2|\beta_2|^2 \nonumber \\
 & & \hspace{-2cm} + (D_{01}\beta_0^*\beta_1^*+D_{02}\beta_0^*\beta_2^*
 +\bar{D}_{12}\beta_1\beta_2^*+ {\rm c.c.})],
\label{2}
\end{eqnarray}
where $\beta_{i}$ ($ i=0,1,2 $) are parameters, c.c. means the
complex conjugated term and
\begin{eqnarray}         %3
 B_0&=&\langle \Delta a_0^{\dagger}\Delta
  a_0\rangle=|f_1|^2+|f_2|^2, \nonumber \\
 B_1&=&\langle \Delta a_1^{\dagger} \Delta a_1\rangle = |g_0|^2, \nonumber \\
 B_2&=& \langle \Delta a_2^{\dagger} \Delta a_2\rangle
  =|h_0|^2, \nonumber  \\
 D_{01}&=&\langle \Delta a_0 \Delta
  a_1\rangle= f_0^*g_0, \nonumber \\
 D_{02}&=& \langle \Delta a_0 \Delta a_2\rangle=
  f_0^*h_0, \nonumber \\
 \bar{D}_{12}&=& -\langle \Delta a_1^{\dagger} \Delta a_2\rangle =
  -h_0g_0^*.
\label{3}
\end{eqnarray}
The functions $ g_0 $, $ h_0 $, and $ f_0 $ are defined in \cite{Allevi2006}.
In order to study the nonclassical nature of the three-mode state, we introduce the determinants
\begin{eqnarray}   % 4
 K_{12}&=& B_1B_2-|\bar{D}_{12}|^2=0, \nonumber \\
 K_{01}&=&B_0B_1-|D_{01}|^2=-|g_0|^2<0, \nonumber \\
 K_{02}&=&B_0B_2-|D_{02}|^2=-|h_0|^2<0.
\label{4}
\end{eqnarray}
According to Eqs.~(\ref{4}) fields $a_1$ and $a_2$ are classically
correlated whereas correlations between fields $a_0$ and $a_1$
($a_0$ and $a_2$) can lead to nonclassical behavior.

In the experiment, the outputs of the detectors placed on fields
$a_1$ and $a_2$ have been summed with the aim to measure
photocount correlations between the single field $a_0$ and the
compound field formed by fields $a_1$ and $a_2$. These
correlations are important to test the performance of a source of
states entangled in the number of photons and obtained after a
conditional measurement of $ n_0 $ photons in field $a_0$. They
can be derived from the following normal characteristic function:
\begin{eqnarray}           %5
 C(\beta_0,\beta_1,\beta_1) &=&
  \exp[-B_0|\beta_0|^2-B_{12}|\beta_1|^2 \nonumber \\
 &+&(D_{0,12}\beta_0^*\beta_1^*+ {\rm h.c.})],
\label{5}
\end{eqnarray}
where
\begin{eqnarray}             %6
 B_{12}&=&B_1+B_2-\bar{D}_{12}-\bar{D}_{21}=|h_0+g_0|^2, \nonumber \\
 D_{0,12}&=&D_{01}+D_{02}=f_0^*g_0 + f_0^*h_0, \nonumber \\
 K_{0,12}&=&B_0B_{12}-|D_{0,12}|^2=-|g_0|^2-|h_0|^2<0.
\label{6}
\end{eqnarray}
To obtain Eq.~(\ref{5}) we have assumed that $\gamma_1 $ is real.
We note that information about the coupling constants
$\gamma_0 $ and $ \gamma_1 $ can be obtained from the reconstruction of the
photocount distributions provided that the
interaction time $t$ is known.

We assume that each field is composed of $ M $ independent temporal modes \cite{Paleari2004}.
Then the variances of integrated intensities $ W $ can be expressed as
follows:
\begin{eqnarray}      %7
 \langle (\Delta W_0)^2 \rangle &=& MB_0^2, \nonumber \\
 \langle (\Delta W_j)^2 \rangle &=& MB_j^2, j=1,2, \nonumber \\
 \langle \Delta W_0\Delta W_j \rangle &=& M |D_{0j}|^2, j=1,2, \nonumber \\
 \langle \Delta W_1\Delta W_2 \rangle &=& M |\bar{D}_{12}|^2.
\end{eqnarray}
If photocounts belonging to fields $a_1$ and $a_2$ are combined together,
we can write the variances of the integrated intensities $ W $ in the following form:
\begin{eqnarray}        %8
 \langle (\Delta W_{12})^2 \rangle &=& \langle (\Delta W_{1})^2
  \rangle+ \langle (\Delta W_{2})^2 \rangle+2
  \langle \Delta W_1\Delta W_2 \rangle \nonumber \\
 &=& M|B_{12}|^2 =M(B_1^2+B_2^2+2|\bar{D}_{12}|^2) \nonumber \\
 &=& M (B_1+B_2)^2, \nonumber \\
 \langle \Delta W_0\Delta W_{12}
   \rangle &=& M |D_{0,12}|^2 =M|f_0|^2(|g_0|^2+|h_0|^2).
\end{eqnarray}
Real value of $\gamma_1$ has been again assumed.

Note that the above-presented analysis can be generalized to
stimulated processes \cite{Perina2008}. However, single as well as
compound fields have to be stimulated in order to support
nonclassical effects by interference terms. As for the compound
field, stimulation of one of its components (i.e. field $ a_1 $ or
$ a_2 $) is sufficient to observe increased nonclassical effects.
Stimulation of only one field (i.e. field $ a_0 $ or $ a_1 $ or $
a_2 $) results in increased values of noise only.

\section{Photon-number distributions and quasi-distributions
of integrated intensities}

\begin{figure}  % figure 1
% [angle=270,width=1\textwidth]
 \begin{center}
  \resizebox{0.8\hsize}{!}{\includegraphics{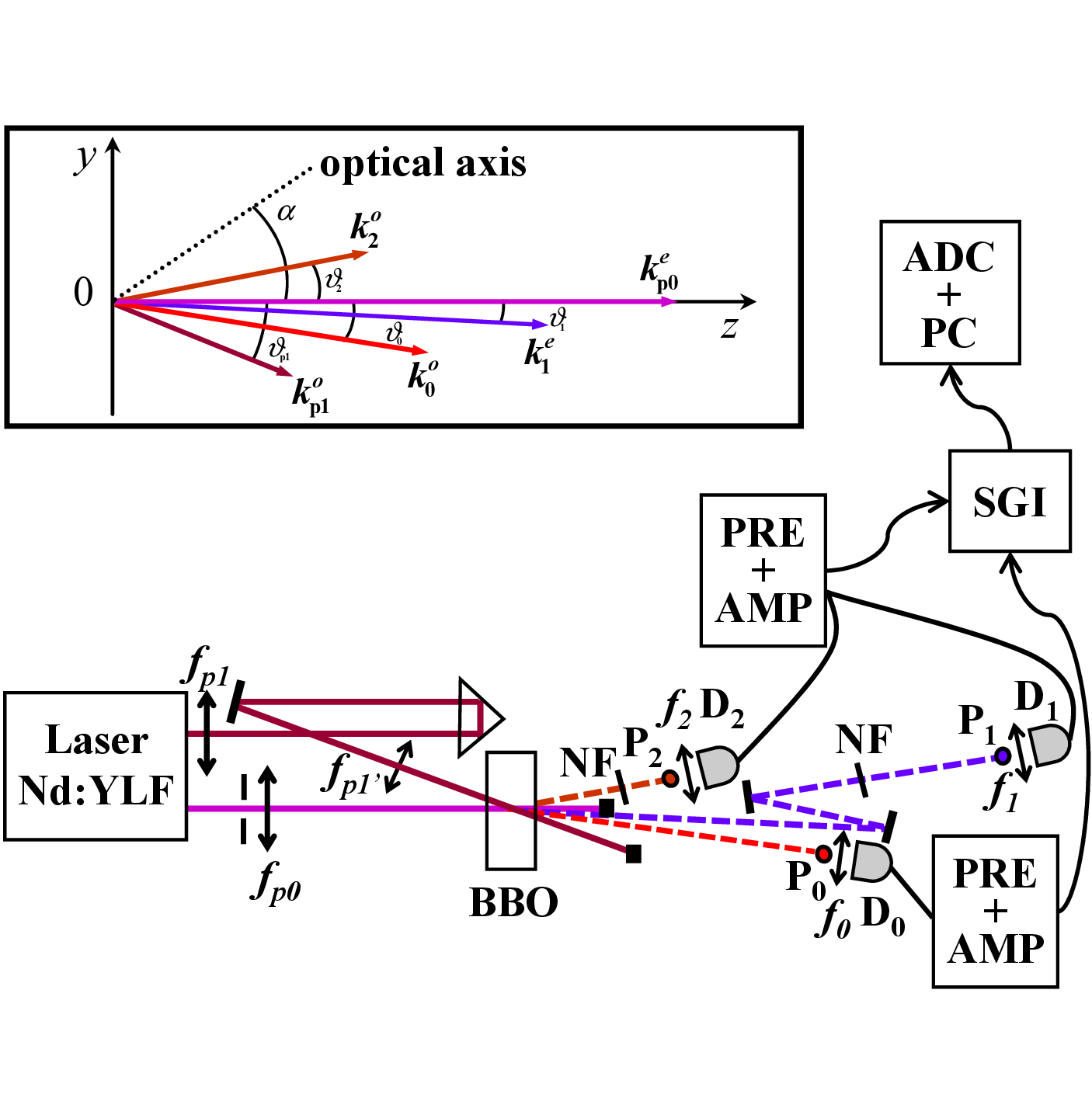}}
 \end{center}
 \caption{Scheme of the experimental setup: BBO, nonlinear crystal;
 NF, variable neutral-density filter; P$_{i}$, pin-holes and D$_{i}$, p-i-n photodiodes,
 $ i=0,1,2 $; $f_{i}$, lenses, $ i=0,1,2,p0,p1,p1' $;
 PRE+AMP, low-noise charge-sensitive pre-amplifiers followed by amplifiers;
 SGI, dual-channel synchronous gated-integrator; ADC+PC, computer integrated
 digitizer.}
\label{fig1}
\end{figure}

The experimental scheme used to generate two nonlinear
interactions is depicted in Fig.~\ref{fig1}.
The harmonics of a continuous-wave mode-locked Nd:YLF laser
regeneratively amplified at a repetition rate of 500 Hz (High Q
Laser Production, Hohenems, Austria) provided two pump fields. In
particular, the third harmonic pulse at 349 nm ($\sim 4.45$ ps
pulse-duration) was exploited as the pump field $a_{p0}$ in
frequency down-conversion, whereas the fundamental pulse at 1047
nm ($\sim 7.7$ ps pulse-duration) was used as the pump field
$a_{p1}$ in parametric amplification. The two interactions
simultaneously satisfied energy-matching
($\omega_{p0}=\omega_{0}+\omega_{2}$ and
$\omega_{1}=\omega_{2}+\omega_{p1}$) and type I phase-matching
(${\mathbf k^e_{p0}}={\mathbf k^o_0}+{\mathbf k^o_2}$, ${\mathbf
k^e_1}={\mathbf k^o_2}+{\mathbf k^o_{p1}}$) conditions, in which
$\omega_{j}$ are the angular frequencies, ${\mathbf k}_j$ denote
the wave vectors and suffixes $o $ and $ e$ indicate ordinary and
extraordinary field polarizations. As depicted in the inset of
Fig.~\ref{fig1}, we set the pump-field $a_{p0}$ direction so that
the wave vector ${\mathbf k_{p0}}$ was normal to the crystal
entrance face and propagated along the $z$-axis of the medium. We
also aligned the wave vector ${\mathbf k_{p1}}$ of the other pump
field $a_{p1}$ in the plane ($y$, $z$) containing the optical axis
(OA) of the crystal and the wave vector ${\mathbf k_{p0}}$. As the
nonlinear medium we used a $\beta$-BaB$_2$O$_4$ crystal (BBO,
Fujian Castech Crystals, China, 10 mm $\times$ 10 mm cross
section, 4 mm thickness) cut for type-I interaction
($\vartheta_{\mathrm{cut}} = 38.4$ deg), into which both pumps
were strongly focused. Typical intensity values of the pump fields
were $\sim 5$ GW/cm$^2$ for $a_{p0}$ and $\sim 2$ GW/cm$^2$ for
$a_{p1}$. The required superposition in time of the two pump
fields was obtained by a variable delay line.

As we have already shown in Ref.~\cite{Allevi2008}, we decided to
generate three fields at non-degenerate frequencies by choosing a
phase-matching condition in the plane ($y$, $z$)
\cite{Bondani2006}. In order to investigate the nature of the
state obtained by the interlinked interactions, we selected a
triplet of coherence areas by means of pin-holes, whose sizes and
distances from the crystal were chosen by searching for the
condition of maximum intensity correlations between the generated
fields \cite{Allevi2008a}. In fact, we expect strong correlations
not only between the number of photons in the field $a_0$ and the
sum of the other two fields (compound field), but also singularly
among the numbers of photons in all pairs of fields. By applying
this criterion, we put two pin-holes of 30 $\mu$m diameter at
distances $d_0 = 60$ cm and $d_2 = 49$ cm from the BBO along the
path of the signal beam at 632.8 nm and of the idler beam at 778.2
nm, respectively. Moreover, as the beam at 446.4 nm has smaller
divergence compared to the other two fields, we selected it by
means of a 50 $\mu$m diameter pin-hole placed at a distance $d_1 =
141.5$ cm from the crystal. The light was suitably filtered by
means of bandpass filters and focused on each detector. In
particular, as we performed measurements in the macroscopic
intensity regime (more than 1000 photons per coherence area), we
used three p-i-n photodiodes (two, D$_{0}$ and D$_{1}$ in
Fig.~\ref{fig1}, S5973-02 and one, D$_2$, S3883, Hamamatsu, Japan)
as the detectors. We obtained the same overall detection
efficiency ($\eta=0.28$) on the three arms by inserting two
adjustable neutral-density filters in the pathways of fields $a_1$
and $a_2$. The current output of the detectors was amplified by
means of two low-noise charge-sensitive pre-amplifiers (CR-110,
Cremat, Watertown, MA) followed by two amplifiers (CR-200-4
$\mu$s, Cremat): to this aim, we connected the detectors $D_1$ and
$D_2$ to the same amplifier device by means of a T-adapter. The
two amplified outputs were then integrated by synchronous
gated-integrators (SGI in Fig.~\ref{fig1}, SR250, Stanford
Research Systems, Palo Alto, CA) sampled, digitized by a 12-bit
converter (AT-MIO-16E-1, DAQ National Instruments) and recorded by
a computer.

From the collected experimental data we obtained the first ($
\langle m_{0} \rangle $, $ \langle m_{12} \rangle $) and the
second ($ \langle m_{0}^2 \rangle $, $ \langle m_{12}^2 \rangle $)
moments of the photocount (photoelectron) distribution of the
single and compound fields, respectively. Moreover, the
correlation between the number of photoelectrons in the single
field and that in the compound field was measured. The unavoidable
contributions of additive noise present in the detection chain can
be quantified in an independent measurement and then subtracted
from the experimental data. By correcting the moments of the
photoelectron distribution for the quantum detection efficiency,
we can obtain the moments for photons. Here we present the first
($\langle n_{0} \rangle $, $ \langle n_{12} \rangle $) and the
second ( $ \langle n_{0}^2 \rangle $, $ \langle n_{12}^2 \rangle
$, $ \langle n_{0}n_{12} \rangle $) moments of the photon-number
distribution:
\begin{eqnarray}            %9
 \langle n_{i} \rangle &=& \langle m_{i} \rangle / \eta , \nonumber \\
 \langle n_{i}^2 \rangle &=& \langle m_{i}^2 \rangle / \eta^2 -
  (1-\eta)\langle m_i \rangle / \eta^2,\  i=0, 12,  \nonumber \\
 \langle n_0 n_{12} \rangle &=& \langle m_0 m_{12} \rangle / \eta^2 .
\end{eqnarray}

In addition, the moments for the integrated intensities can be
determined as follows:
\begin{eqnarray}    %10
 \langle W_i \rangle &=& \langle n_i \rangle , \nonumber \\
 \langle W_i^2 \rangle &=& \langle n_i^2 \rangle - \langle n_i \rangle,
  i=0, 12,  \nonumber \\
 \langle W_0 W_{12} \rangle &=& \langle n_0 n_{12} \rangle .
\end{eqnarray}
In a theory based on the generalized superposition of signal and
noise properties of the fields can be quantified using
coefficients $B_0 $, $ B_{12} $, $ D_{0,12} $ and number $ M $ of
temporal modes. These quantities can be determined from the
moments of integrated intensities \cite{Perina2005}:
\begin{eqnarray}   % 11
 B_{i} &=& \langle (\Delta W_{i})^2 \rangle/\langle W_{i} \rangle ,
  \nonumber \\
 M_i &=& \langle W_{i} \rangle^2/\langle (\Delta W_{i})^2 \rangle,
  \hspace{2mm} i=0,12 , \nonumber \\
 |D_{0,12}| &=& \sqrt{ \langle \Delta W_0 \Delta W_{12} \rangle/M } .
\label{11}
\end{eqnarray}
The mean number of photons in mode $ i $ is given by $ B_i $ ($
i=0,12 $) whereas $ D_{0,12} $ quantifies the correlations between
the single field and the compound one. The number of modes $ M $
can be determined either from the experimental data measured in
the single or in the compound field (see Eqs.~(\ref{11})).
Ideally, $ M_0 $ should be equal to $M_{12} $ \cite{Allevi2006}.
This cannot be reached in a real experiment because of non-perfect
alignment and detection noise. However, a correct alignment of the
experimental setup allows to have $ M_0 \approx M_{12} $ so that
we can define $ M = (M_0 + M_{12})/2 $. A more detailed analysis
concerning the determination of $M$ can be found in
\cite{Perina2007}.

The JSCF photon-number distribution $ p(n_0,n_{12})$, which can be
derived from the normal characteristic function $ C $ in
Eq.~(\ref{5}), is written as follows \cite{Perina2005}:
\begin{eqnarray}   % 12
 p(n_0,n_{12}) &=& \frac{1}{\Gamma(M)} \frac{(B_0+K_{0,12})^{n_0}
(B_{12}+K_{0,12})^{n_{12}}}{(1+B_0+B_{12}+K_{0,12})^{n_0+n_{12}+M}}
\nonumber \\
 & & \hspace{-1cm} \mbox{} \times
  \sum_{r=0}^{{\rm min}(n_0,n_{12})} \frac{\Gamma(n_0+n_{12}+M-r)}
{r!  (n_{0}-r)!   (n_{12}-r)!} \nonumber \\
 & & \hspace{-1cm} \mbox{} \times  \frac{(-K_{0,12})^r
  (1+B_0+B_{12}+K_{0,12})^r}{ [(B_0+K_{0,12})(B_{12}+K_{0,12})]^{r}},
\label{12}
\end{eqnarray}
where $\Gamma$ is the gamma function. If determinant $ K_{0,12}$
defined in Eqs.~(\ref{6}) is negative, the given field cannot be
described classically. The quantities $ B_0 + K_{0,12} $ and $
B_{12} + K_{0,12} $ occurring in Eq.~(\ref{12}) cannot be negative
and can be interpreted as components of fictitious noise. In the
ideal lossless case, $ K_{0,12} = -B_0 =-B_{12} $ and then the
joint photon-number distribution $ p(n_0,n_{12}) $ takes the form
of the diagonal Mandel-Rice distribution. As frequency
down-conversion produces couples of photons, we expect that the
highest values of the elements of $ p(n_0,n_{12}) $ are near the
diagonal $ n_0 = n_{12} $. In addition certain classical
inequalities can be violated in this region \cite{Haderka2005}.

If $ K_{0,12}=0 $, $i.e.$ at the boundary between the classical
and nonclassical behaviors, the compound Mandel-Rice formula for
JSCF photon-number distribution $ p(n_0,n_{12})$ can be simplified
as follows
\begin{equation}   %13
 p(n_0,n_{12})= \frac{ \Gamma(n_0+n_{12}+M)B_0^{n_0}B_{12}^{n_{12}} }{
 \Gamma(M)n_0!\, n_{12}! \, (1+B_0+B_{12})^{n_0+n_{12}+M} }.
\label{13}
\end{equation}

In the ideal case only the diagonal elements of the distribution
$p(n_0,n_{12})$ are different from zero: in this case the
post-selection scheme for the preparation of a state entangled in
the number of photons perfectly works. However, losses present in
any experimental implementation cause discrepancy from this ideal
situation. In order to quantify this discrepancy, we determine the
conditional compound-field photon-number distribution $
p_{c,12}(n_{12};n_0) $ provided that $ n_0 $ photons are detected
in the single field:
\begin{equation}    % 14
 p_{c,12}(n_{12};n_0)= p(n_0,n_{12})/\sum_{k=0}^{\infty}p(n_0,k) .
\label{14}
\end{equation}
Fano factor $ F_{c,12} $ of photon-number distribution $
p_{c,12}(n_{12};n_0) $ can be expressed as follows:
\begin{eqnarray}   % 15
 F_{c,12}(n_0) &=& 1 \nonumber \\
 & & \hspace{-2cm} \mbox{} + \frac{(1+M/n_0)[(B_{12}+K_{0,12})/(1+B_0)]^2
 -(K_{0,12}/B_0)^2}{ (1+M/n_0)(B_{12}+K_{0,12})/(1+B_0)- K_{0,12}/B_0}  \nonumber \\
 & \approx &  1+K_{0,12}/B_0.
\label{15}
\end{eqnarray}
Note that the last approximation in Eq.~(\ref{15}) holds for
$K_{0,12} \approx -B_{12} $ and highlights that negative values of
determinant $ K_{0,12} $ cause sub-Poissonian conditional
photon-number distribution. For the ideal lossless case, $
K_{0,12}=-B_0 =-B_{12} $ and Fano factor $ F_{c,12} $ is equal to
0.

The correlations in the number of photons can also be quantified by
the distribution $ p_{-} $ of the difference photon
number $ n_0-n_1-n_2 $ defined as:
\begin{equation} % 16
 p_{-}(n) = \sum_{n_0,n_1,n_2=0}^{\infty} \delta_{n,n_0-n_1-n_2}
  p(n_0,n_1,n_2),
\label{16}
\end{equation}
where $ \delta $ is the Kronecker symbol. The variance of the
difference $ n_0-n_1-n_2 $ can be lower than the sum $ \langle n_0
+ n_1 + n_2 \rangle $ of the mean photon numbers in all fields.
This happens for nonclassical fields and we obtain sub-shot-noise
correlations in this case \cite{Bondani2007,Allevi2008}. Under
this condition the so-called noise reduction factor
\cite{Bondani2007},
\begin{equation}   % 17
 R= \frac{ \langle [\Delta(n_0-n_1-n_2)]^2 \rangle }{\langle n_0 \rangle +
 \langle n_1 \rangle + \langle n_2 \rangle} ,
\label{17}
\end{equation}
is lower than 1.

The JSCF photon-number distribution $ p(n_0,n_{12}) $ is given by
Mandel's photo-detection formula
\cite{Perina1994,Saleh1978,Perina2005}, which is defined in terms
of the $ s $-ordered quasi-distribution $P_s(W_0,W_{12}) $ of
integrated intensities. This formula can be inverted in such a way
that the quasi-distributions of integrated intensities can be
written in terms of the distribution $ p(n_0,n_{12}) $ in
Eq.~(\ref{12}) with the following results. If we have the $ s
$-ordered determinant $ K_{0,12s}
> 0 $ ($ K_{0,12s}=B_{0s}B_{12s}-|D_{0,12}|^2 $, $ B_{i,s}=B_{i}+(1-s)/2
$, $ i=0,12 $), the $ s $-ordered JSCF quasi-distribution $
P_s(W_0,W_{12}) $ of integrated intensities is a non-negative
ordinary function \cite{Perina2005}:
\begin{eqnarray}  % 18
 P_s(W_0,W_{12}) &=&  \frac{1}{\Gamma(M) K_{0,12s}^M}
  \left(\frac{K_{0,12s}^2 W_0W_{12}}{|D_{0,12}|^2}\right)^{(M-1)/2}
  \nonumber \\
 & & \hspace{0cm} \mbox{} \times \exp\left[-\frac{(B_{12s}W_0/B_{0s}+W_{12})
  B_{0s}}{K_{0,12s}}\right]
  \nonumber \\
 & & \mbox{} \times  I_{M-1}\left(2
  \sqrt{\frac{|D_{0,12}|^2W_0W_{12}}{K_{0,12s}^2} }\right) ,
\label{18}
\end{eqnarray}
where $ I_{M} $ is the modified Bessel function.

On the other hand, for $ K_{0,12s} < 0 $, the JSCF
quasi-distribution $ P_s(W_0,W_{12}) $ of integrated intensities
takes the form of a generalized function. It can be approximated
using the following expression \cite{Perina2005}:
\begin{eqnarray}   % 19
 P_s(W_0,W_{12}) & \approx & \frac{A(W_0W_{12})^{(M-1)/2}}{\pi \Gamma(M)
(B_{0s}B_{12s})^{M/2}} \nonumber \\
 & & \mbox{} \hspace{-2cm} \times
  \exp\left(-\frac{W_0}{2B_{0s}} -\frac{W_{12}}{2B_{12s} }\right) \nonumber \\
 & & \mbox{} \hspace{-2cm} \times
  \rm{sinc} \left[ A\left(\sqrt{\frac{B_{12s}}{B_{0s}} } W_0-
   \sqrt{ \frac{B_{0s}}{B_{12s}} }  W_{12}\right) \right],
\label{19}
\end{eqnarray}
in which $ {\rm sinc}(x) = \sin(x)/x $ and
$A=(-K_{0,12s})^{-1/2}$. The quasi-distribution derived in
Eq.~(\ref{19}) typically oscillates and has negative values in
some regions. The threshold value $ s_{\rm th} $ giving the
boundary between the expressions in Eqs. (\ref{18}) and (\ref{19})
can be determined from the condition $K_{0,12s}=0$:
\begin{equation}  % 20
 s_{\rm th} = 1 + B_0+B_{12} - \sqrt{(B_0+B_{12})^2 -
  4 K_{0,12}},
\label{20}
\end{equation}
in which $ -1 \le s_{\rm th} \le 1 $.

The variances of the difference $ n_0 - n_{12} $ of single- and
compound-field photon numbers ($ n_{12} = n_1 + n_2 $) and
the difference $ W_0 - W_{12} $ of the single- and compound-field
integrated intensities ($ W_{12} = W_1 + W_2 $) are linked by the following formula:
\begin{equation}   % 21
 \langle [\Delta(n_0-n_{12})]^2 \rangle = \langle n_0 \rangle +
  \langle n_{12} \rangle + \langle [\Delta(W_0-W_{12})]^2  \rangle .
\label{21}
\end{equation}
Equation (\ref{21}) shows that negative values of the
quasi-distribution $ P_s(W_0,W_{12}) $ of integrated intensities
are needed to observe sub-shot-noise correlations in single- and
compound-field photon numbers, i.e. $ R < 1 $ in Eq.~(\ref{17}).

\section{Experimental results and their interpretation}

From the intensity measurements of the single and compound fields
we calculated the following moments for photoelectrons: $ \langle
m_0 \rangle = 1225.183 $, $\langle m_{12} \rangle = 1186.138 $, $
\langle m_0^2 \rangle = 1609827 $, $ \langle m_{12}^2 \rangle =
1518257 $, $ \langle m_0 m_{12} \rangle = 1562402 $. Moreover, the
knowledge of the quantum detection efficiencies $ \eta_0 = \eta_1
= \eta_2 = 0.28 $ allowed us to derive the corresponding moments
for photons: $ \langle n_0 \rangle = 4375.654 $, $ \langle n_{12}
\rangle = 4236209 $, $ \langle n_0^2 \rangle = 20522260 $, $
\langle n_{12}^2 \rangle = 19354630 $, $ \langle n_0 n_{12}
\rangle = 19928600 $.

In the case of photoelectrons the calculated values of the
coefficients in Eqs.~(\ref{11}) are:
\begin{eqnarray}   % 22
 B_0=87.765104,  B_{12}=92.861384 , \nonumber \\
 |D_{0,12}|= 90.371968, M=13.3665.
\label{22}
\end{eqnarray}
We also note that the number of modes $M_0=13.95980 $ and $
M_{12}=12.77321 $ relative to the single and compound fields,
respectively, are almost the same. In the case of photons the
values of coefficients $ B_0 $, $ B_{12} $ and $ |D_{0,12}| $ are
the following:
\begin{eqnarray}   % 23
 & & B_0=313.447,  B_{12}=331.648 , \nonumber \\
 & & |D_{0,12}|= 322.757.
\label{23}
\end{eqnarray}
The number of modes $ M $ is the same for photocount and photon-number
distributions.

Determinant $ K_{0,12} $ is negative both for photoelectron- ($
K_{0,12}= -17.104 $) and photon-number ($ K_{0,12} = -218.158 $)
distributions. At the same time, the fluctuations in the
difference between the single and the compound fields both for
photoelectrons ($ R = 0.954 $) and for photons ($ R = 0.837 $) are
below the shot-noise level. In accordance with Eq.~(\ref{21}),
this means that the wave variance $ \langle [\Delta
(W_0-W_{12})]^2 \rangle $ of the difference between the single and
compound field integrated intensities is negative: in fact we have
the wave variance equal to $ - 110.572 $ for photoelectrons and
$-1406.699$ for photons. These negative values are due to the fact
that frequency down-conversion emits the same number of photons
into the single field $a_0$ and in the compound field formed by
fields $a_1$ and $a_2$. For the same reason we obtain that the
covariance $ C=\langle \Delta m_0 \Delta m_{12} \rangle/ \sqrt{
\langle (\Delta m_0)^2\rangle \langle(\Delta m_{12})^2 \rangle } $
for photoelectrons and the analogous expression for photons assume
values very close to 1: $0.991 $ for photoelectrons and $0.990$
for photons. Moreover, we note that even the principal squeezing
parameter $\lambda = 1 + B_0 + B_{12} - 2|D_{0,12}|$
\cite{Perina1991} indicates a nonclassical behavior: we obtained
$\lambda = 0.882 $ for photoelectrons and $ 0.581 $ for photons; $
\lambda = 1 $ holds for coherent states. On the basis of the above
mentioned quantities, we want to emphasize that the quantum nature
of the state produced by the nonlinear process is more evident in
terms of photons than in terms of photoelectrons.

\begin{figure} %figure2
  a)

 \begin{center}
  \resizebox{0.8\hsize}{!}{\includegraphics{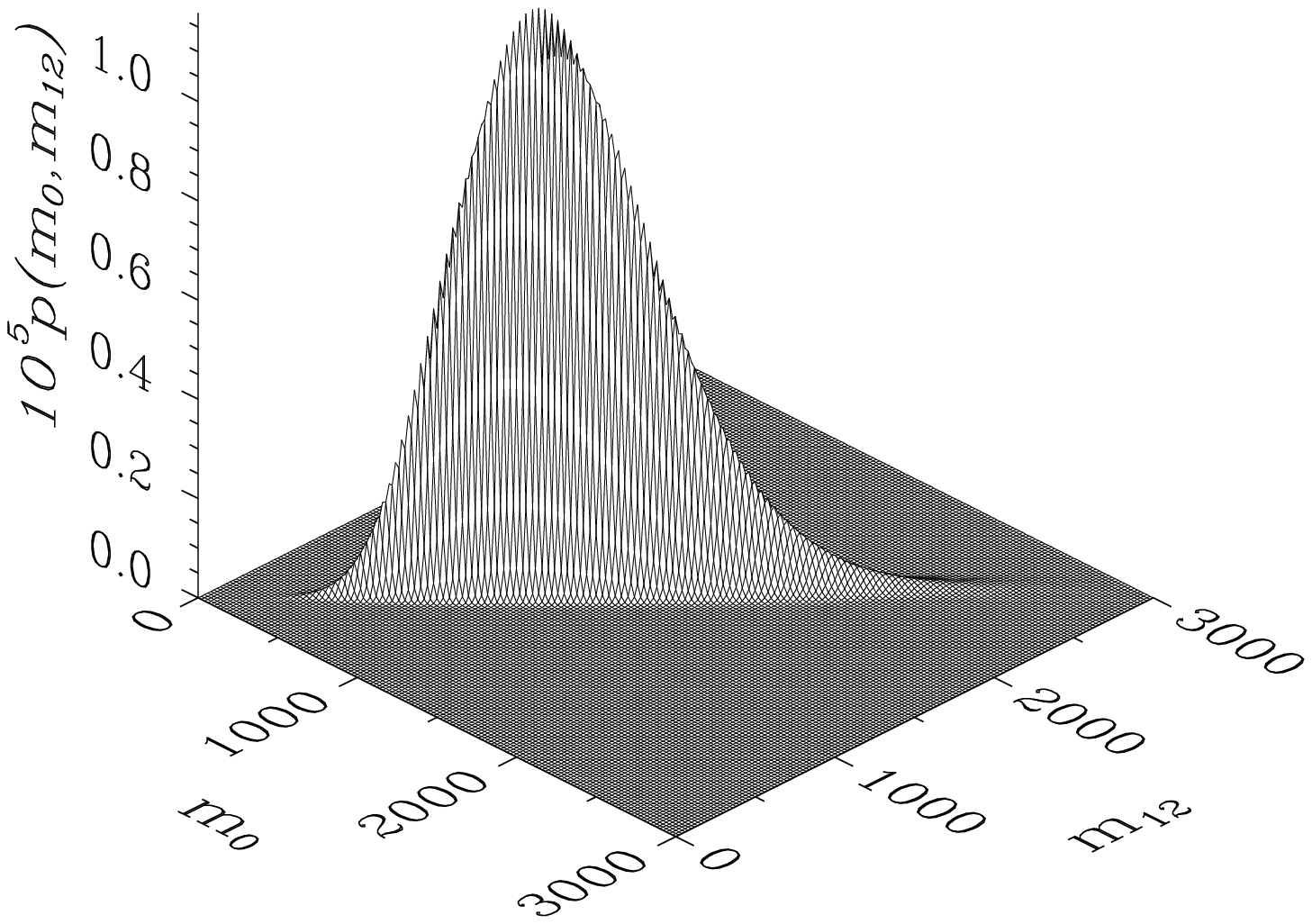}}
 \end{center}
  b)

 \begin{center}
  \resizebox{0.6\hsize}{!}{\includegraphics{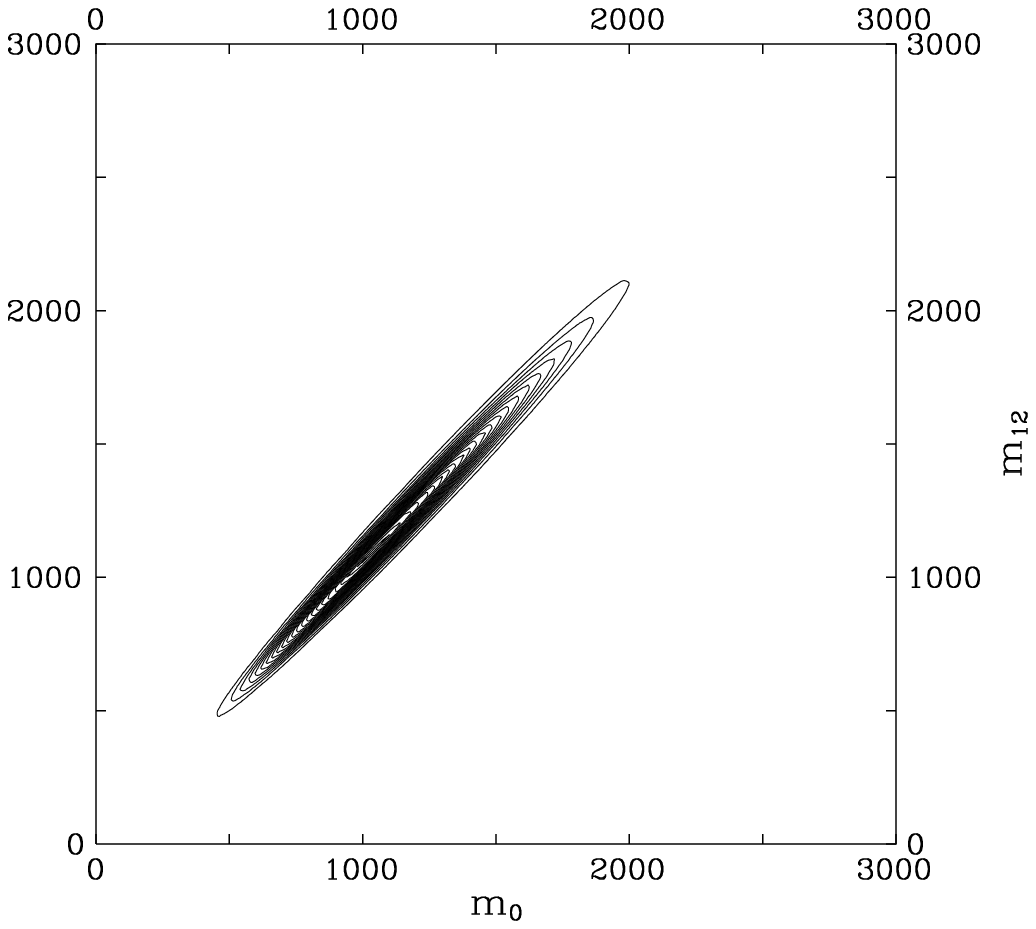}}
 \end{center}
  \caption{JSCF photocount distribution
  $p(m_0,m_{12})$ (a) and its contour plot (b)}
\label{fig2}
\end{figure}

\begin{figure} %figure3
  a)

 \begin{center}
  \resizebox{0.8\hsize}{!}{\includegraphics{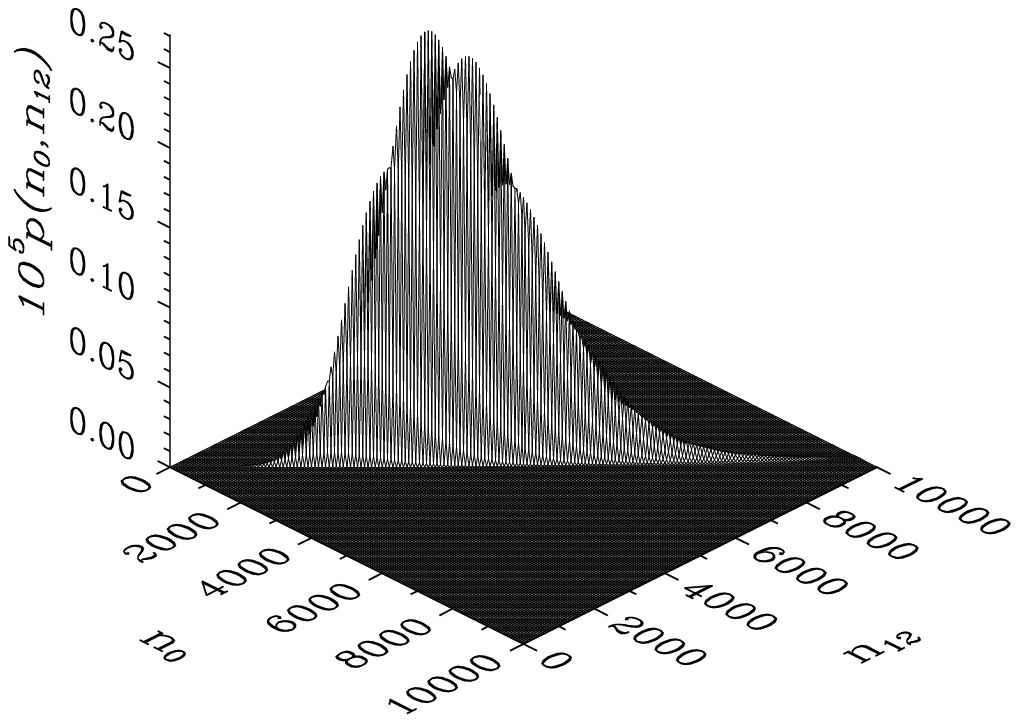}}
 \end{center}

  b)

 \begin{center}
  \resizebox{0.6\hsize}{!}{\includegraphics{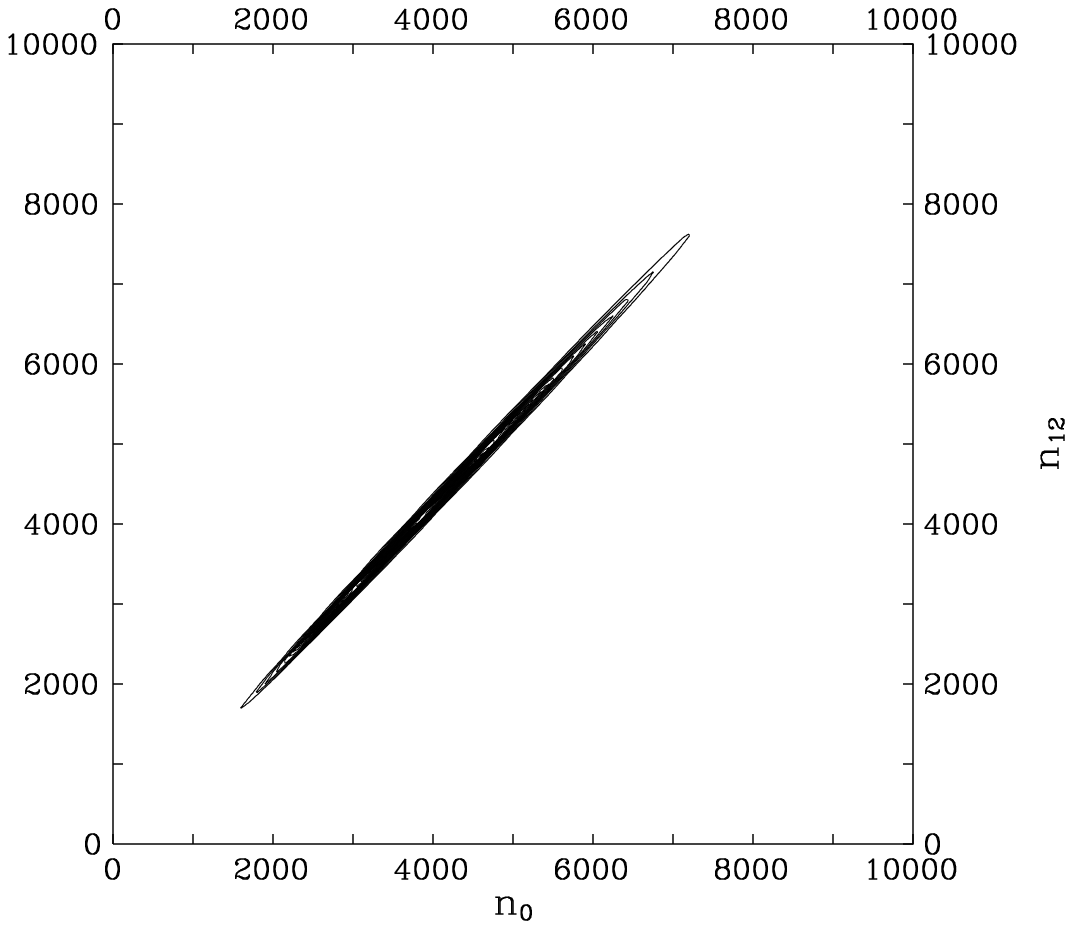}}
 \end{center}

  \caption{JSCF photon-number distribution $
  p(n_0,n_{12})$ (a) and its contour plot (b).}
\label{fig3}
\end{figure}

The JSCF photocount, $ p(m_0,m_{12}) $, and photon-number, $
p(n_0,n_{12}) $, distributions calculated along the formula in
Eq.~(\ref{12}) with values of parameters appearing in
Eqs.~(\ref{22}) and (\ref{23}) are shown in Fig.~\ref{fig2} and
Fig.~\ref{fig3}, respectively. There are strong photon-number
correlations between the single field and the compound one. In
particular, we note that the elements of $ p(m_0,m_{12}) $ and $
p(n_0,n_{12}) $ assume the highest values in the vicinity of the
diagonal, $i.e.$ for $m_0 \approx m_{12} $ and $n_0 \approx n_{12}
$ (see contour plots in Figs.~\ref{fig2} and \ref{fig3}).

\begin{figure}    %figure4
 \begin{center}
  \resizebox{0.6\hsize}{!}{\includegraphics{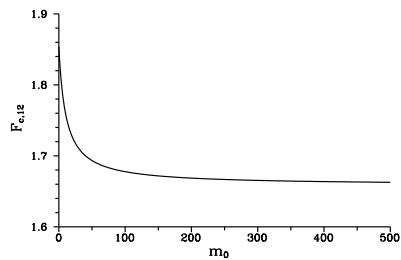}}
 \end{center}
  \caption{Fano factor $ F_{c,12} $ of the compound-field
  conditional photocount distribution $ p_{c,12} $ as a function of
  the number $ m_0 $ of detected single-field photoelectrons.}
\label{fig4}
\end{figure}

\begin{figure}    %figure5
 \begin{center}
  \resizebox{0.6\hsize}{!}{\includegraphics{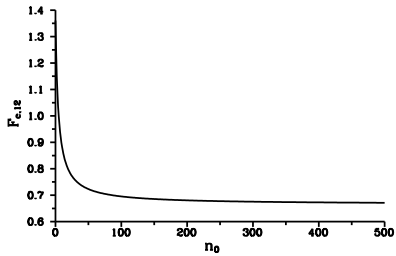}}
 \end{center}
 \caption{Fano factor $F_{c,12}$ of the compound-field conditional
  photon-number distribution $p_{c,12}$ as a function of the number
  $n_0$ of detected single-field photons.}
\label{fig5}
\end{figure}

Values of Fano factors $F_{c,12}$ of the conditional photocount
and photon-number distributions are important from the point of
view of the conditional state-preparation scheme. We remind that a
nonclassical state requires $ F_{c,12} < 1 $, i.e. sub-Poissonian
statistics of the conditional distributions. In our case this
requirement is not fulfilled by the photocount distribution, as
shown in Fig.~\ref{fig4}. This behavior is due to the relatively
low quantum detection efficiency ($ \eta = 0.28 $). On the other
hand, if the experimental data are corrected for the quantum
efficiency, we can obtain a sub-Poissonian photon-number
distribution for $ n_0 > 5 $ (see Fig.~\ref{fig5}). The greater
the number $ n_0 $ of detected single-field photons the smaller
the value of Fano factor $ F_{c,12} $. Actually, the values of
Fano factor $ F_{c,12} $ do not change for numbers of $ m_0 $ and
$ n_0 $ greater than 100.

\begin{figure} %figure6
  a)

 \begin{center}
  \resizebox{0.8\hsize}{!}{\includegraphics{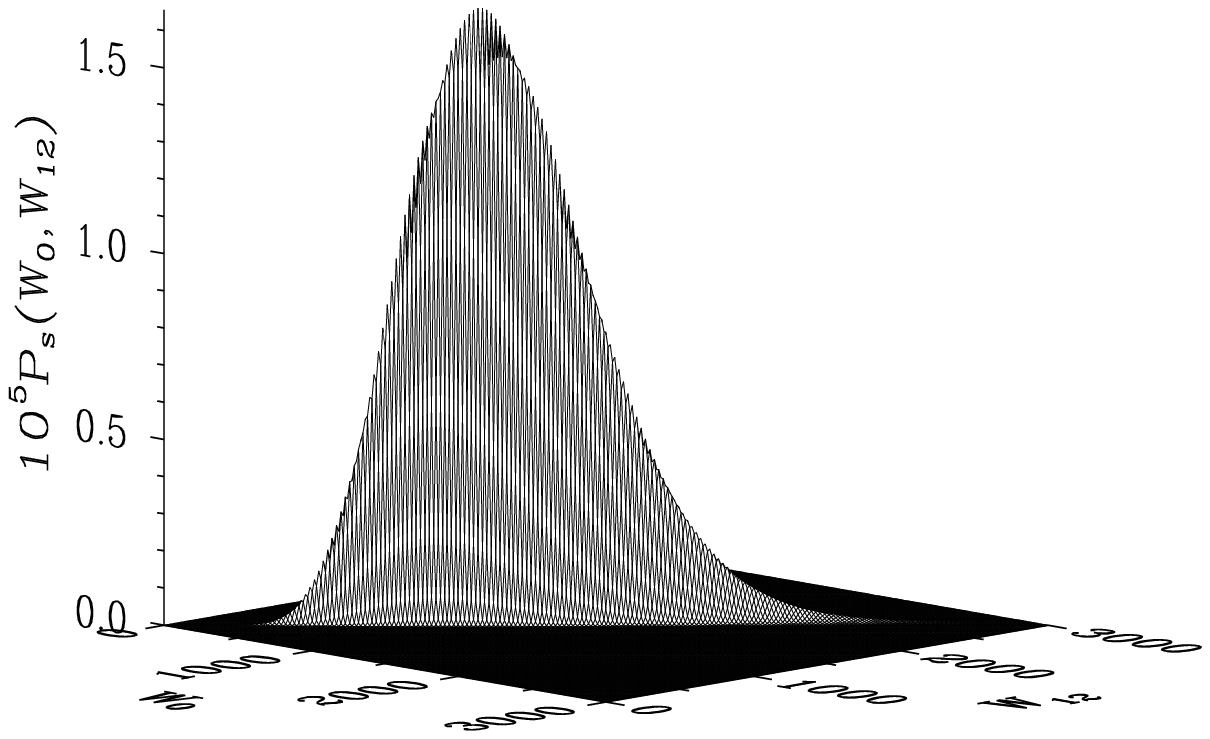}}
 \end{center}

  b)

 \begin{center}
  \resizebox{0.8\hsize}{!}{\includegraphics{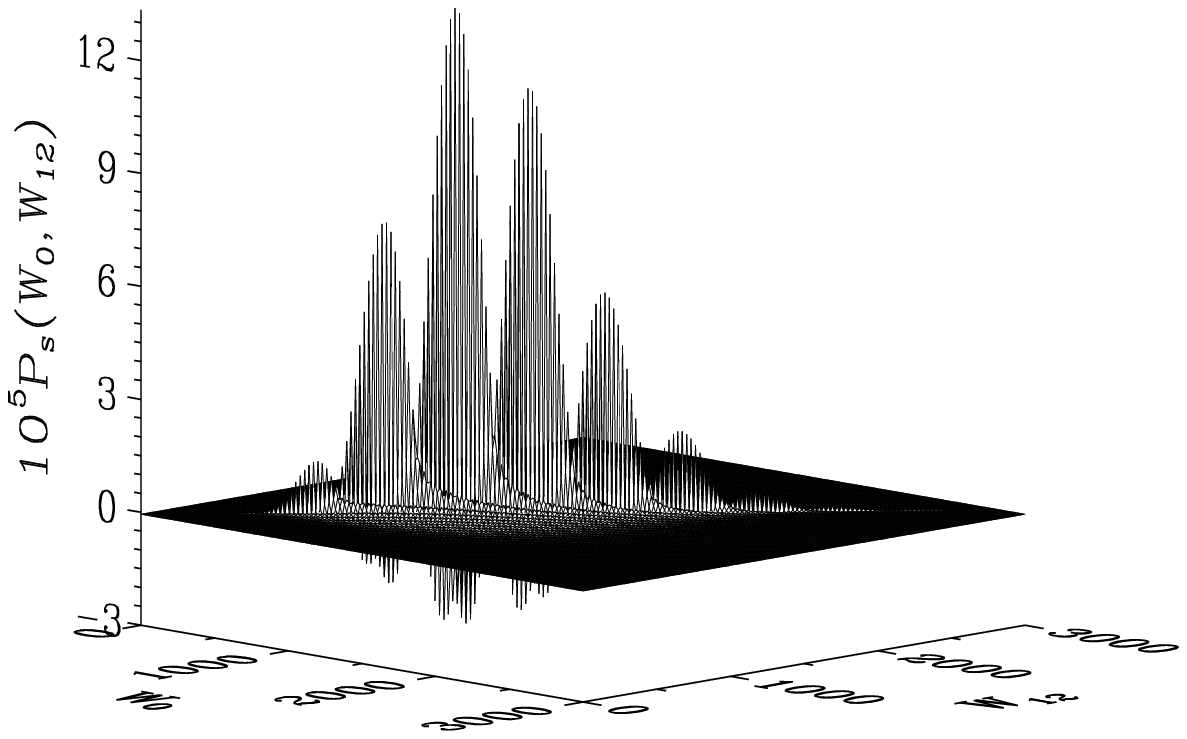}}
 \end{center}

  \caption{JSCF quasi-distributions $P_s(W_0,W_{12})$ of the single-field ($W_0$)
  and compound-field ($W_{12}$) integrated intensities corresponding to photoelectrons
  for $s=0.4$ (a) and $s=0.9$ (b).}
\label{fig6}
\end{figure}

\begin{figure} %figure7
  a)

 \begin{center}
  \resizebox{0.8\hsize}{!}{\includegraphics{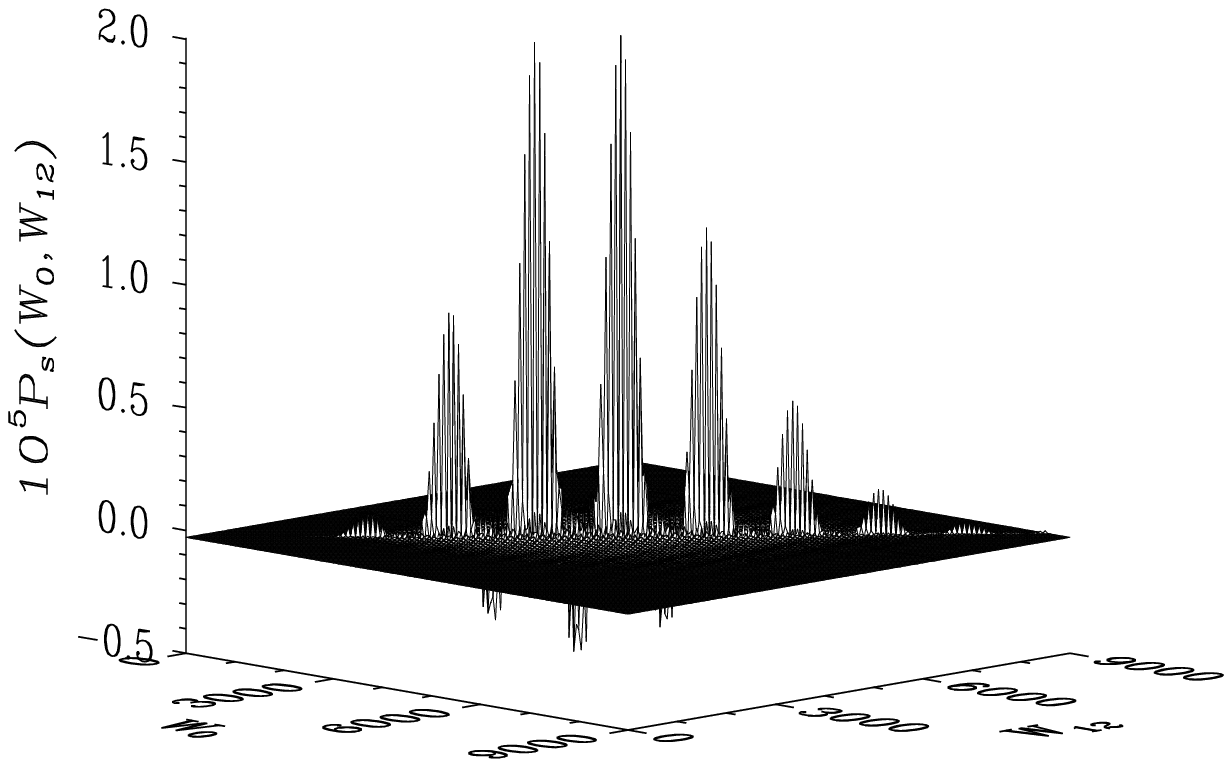}}
 \end{center}

  b)

 \begin{center}
  \resizebox{0.6\hsize}{!}{\includegraphics{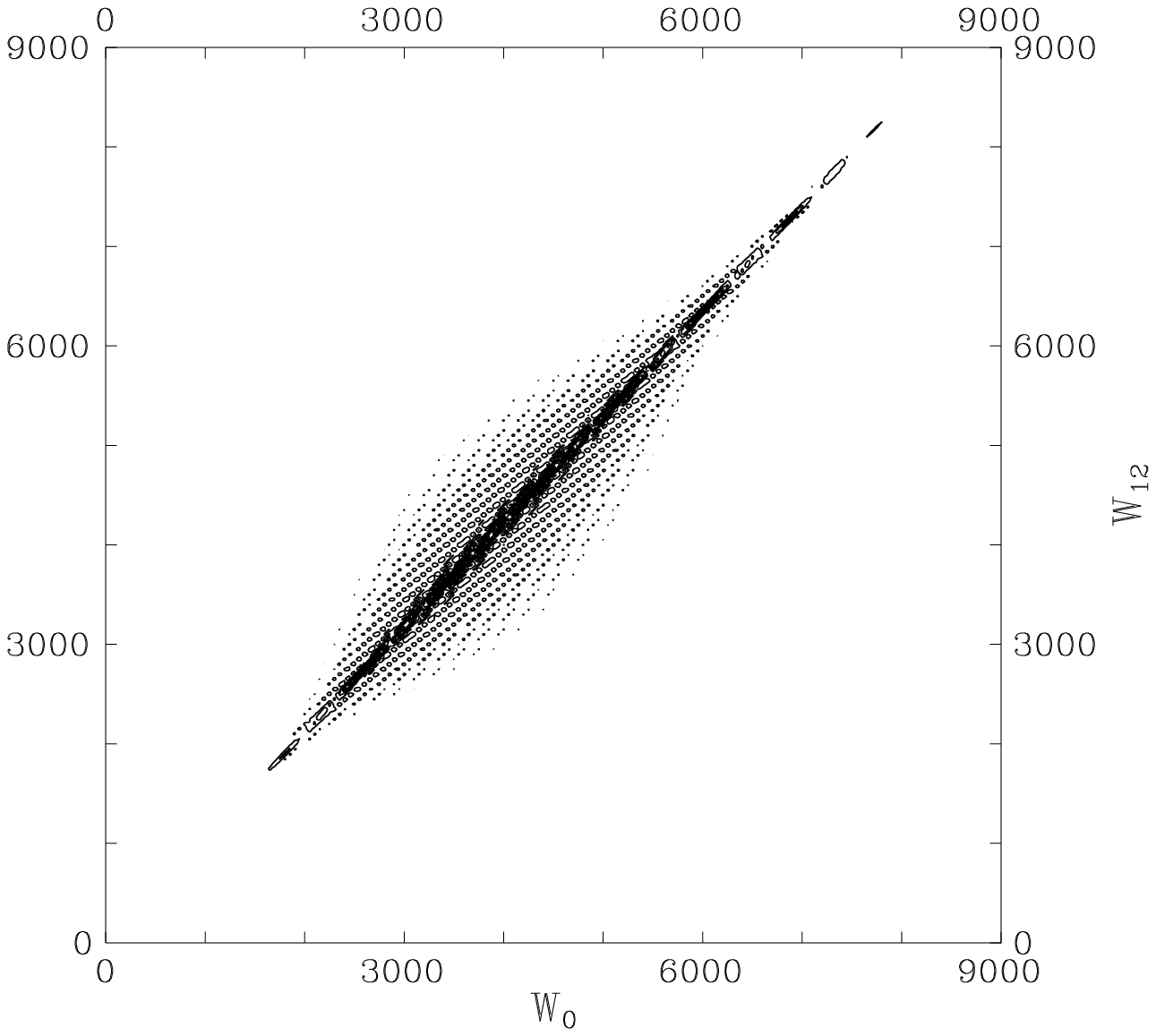}}
 \end{center}

  \caption{JSCF quasi-distribution $P_s(W_0,W_{12})$ of the single-field ($W_0$) and compound-field
  ($W_{12}$) integrated intensities of photons for $s=0.4$ (a) and the corresponding contour plot (b).}
\label{fig7}
\end{figure}

The behavior of the JSCF quasi-distribution $P_s(W_0,W_{12})$ of
integrated intensities depends on the value of the ordering
parameter $ s $. In our case the threshold value of the ordering
parameter, $s_{\rm th} $, is equal to 0.811 for photoelectrons.
This means that the nonclassical behavior of the
quasi-distribution $P_s(W_0,W_{12})$ is expected only for values
of $ s $ greater than $ s_{\rm th} $. Indeed, in Fig.~\ref{fig6}
we show that oscillations and negative values of $P_s(W_0,W_{12})$
occur for $s=0.9 $ ($K_{0,12s}<0$); on the other hand, the
nonclassical features do not appear for $ s=0.4 $ ($K_{0,12s}>0$).
This means that the quantum noise present in the detection chain
covers the nonclassical behavior. In the case of photons, the JSCF
quasi-distribution $P_s(W_0,W_{12})$ has a lower threshold value
of the ordering parameter, namely $ s_{\rm th} =0.324$. For this
reason, the JSCF quasi-distribution $P_s(W_0,W_{12})$ shown in
Fig.~\ref{fig7} already shows nonclassical features like
oscillations and negative values for $ s=0.4 $.

\begin{figure} %figure8
  a)

 \begin{center}
  \resizebox{0.6\hsize}{!}{\includegraphics{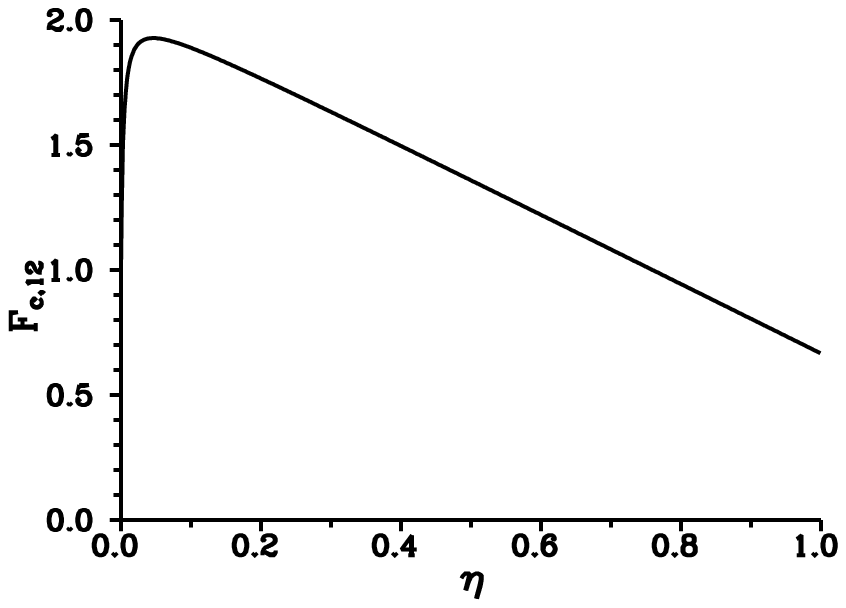}}
 \end{center}

  b)

 \begin{center}
  \resizebox{0.6\hsize}{!}{\includegraphics{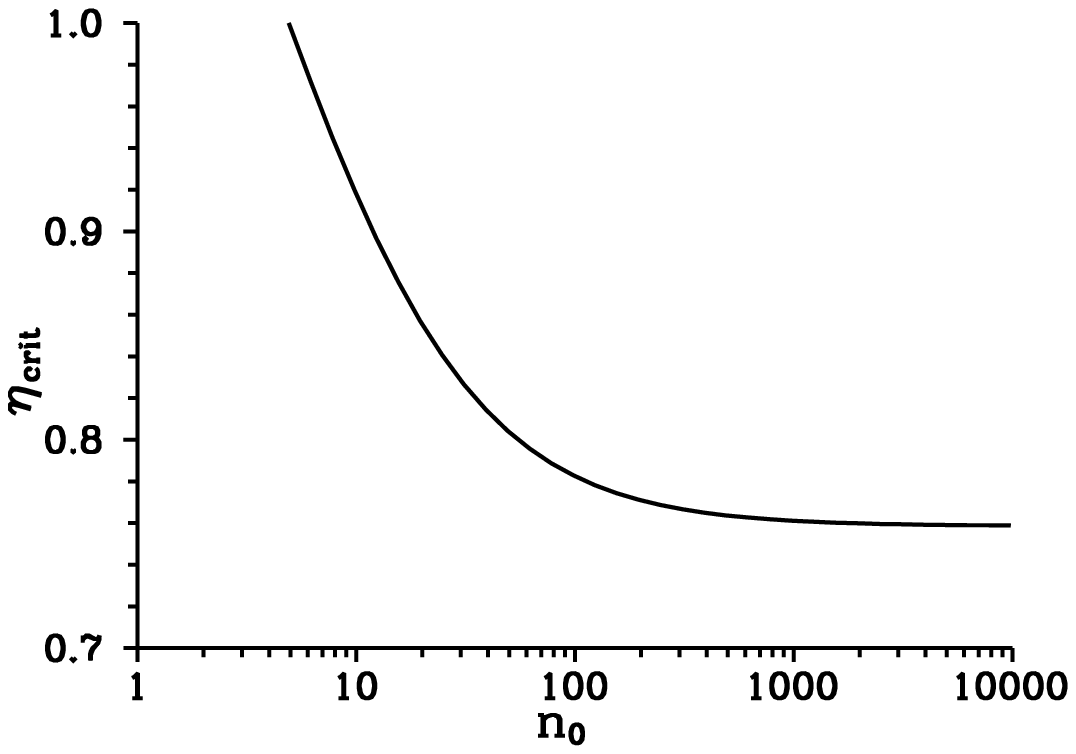}}
 \end{center}

  \caption{Conditional Fano factor $F_{c,12}$ as a function of the quantum efficiency
  $\eta=\eta_0=\eta_1=\eta_2$ for the fixed number $n_0=3000$ of the single-field photons
  (a) and critical quantum detection efficiency $\eta_{\rm crit}$ as a function of the number $ n_0 $
  of the single-field photons (b). If $\eta>\eta_{\rm crit}$ then $F_{c,12} <1$ and viceversa.
  In graph (b), logarithmic scale is used on the $ x $-axis.}
\label{fig8}
\end{figure}

\begin{figure} %figure9
\begin{center}
 \resizebox{0.8\hsize}{!}{\includegraphics{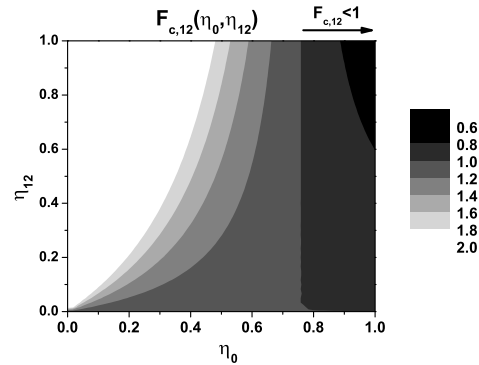}}
\end{center}
  \caption{Contour plot of the conditional Fano factor $ F_{c,12}$ as a function of the quantum efficiencies
  $\eta_0 $ and $ \eta_{12}$, $ \eta_{12}=\eta_1=\eta_2 $ for
  high values of $n_0$ ($ n_0 \rightarrow \infty $).
  The values in the white area in the upper left corner have not been determined.
  $ F_{c,12} < 1 $ for $ \eta_0 > \eta_{0,\rm crit,\rm min} $,
  $ \eta_{0,\rm crit,\rm min}=0.7585 $.}
\label{fig9}
\end{figure}

It is clear from all these considerations that to appreciate the
quantum features of the three-mode state and its usefulness for
the preparation of a conditional state, the quantum detection
efficiencies must be high enough. A typical dependence of the
conditional Fano factor $ F_{c,12}(n_0) $ on the quantum detection
efficiency $ \eta $ ($ \eta = \eta_0 = \eta_1 = \eta_2 $) is
plotted in Fig.~\ref{fig8} a) for $ n_0 = 3000 $. By decreasing
the value of $ \eta $ we obtain an increase in the value of
$F_{c,12}(n_0) $. Moreover, it can be demonstrated that there is a
threshold value of the efficiency $\eta_{\rm crit} $ below which
the conditional compound-field photon-number distribution is no
more sub-Poissonian. As shown in Fig.~\ref{fig8} b), the critical
value $ \eta_{\rm crit} $ decreases by increasing the number $ n_0
$ of the photons in the single field. We note that graphs in
Figs.~\ref{fig8} and \ref{fig9} have been obtained using the
formula for Fano factor $ F_{c,12} $ in Eq.~(\ref{15}) assuming
substitution $ B_0 \rightarrow \eta_0 B_0 $, $ B_{12} \rightarrow
\eta_1 B_{12} $, and $ D_{0,12}^2 \rightarrow \eta_0 \eta_1
D_{0,12}^2 $ and coefficients $ B_0 $, $ B_{12} $, and $ D_{0,12}
$ for photons [Eqs.~(\ref{23})]

In more detail, only the value of the quantum detection efficiency
$ \eta_0 $ is crucial for the sub-Poissonian behavior of the
conditional compound-field photon-number distribution, as shown in
Fig.~\ref{fig9}, where the contour plot of Fano factor $F_{c,12}$
is depicted as a function of detection efficiencies $\eta_0$ and
$\eta_{12}$ for $n_0\rightarrow\infty$. In fact, for $ \eta_0 >
\eta_{0,\rm crit, min}=0.7585$ we have a sub-Poissonian behavior,
which does not depend on the value of $ \eta_{12} $ (we suppose
that $ \eta_{12} = \eta_1 = \eta_2 $). This means that when
$\eta_0$ exceeds its critical value, the quantum detection
efficiency corresponding to the conditionally prepared state will
not affect the nonclassical behavior of the state itself.

Note that, from the experimental point of view, the nonclassical
behavior of an optical state is usually exhibited by the condition
$ R <1 $. On the other hand, in accordance with the theoretical
study presented in \cite{Perina2005}, nonclassical fields satisfy
the condition $ K_{0,12} < 0 $. Actually, the two conditions can
be linked together. In fact, if $ B_0 = B_{12} = B $, negative
values of the determinant $ K_{0,12} $, which testify a
nonclassical character, lead to
$\langle[\Delta(W_0-W_{12})]^2\rangle = 2M(B^2-|D_{0,12}|^2)<0 $.
By using Eq.~(\ref{17}) we thus obtain that $ R < 1 $, i.e. the
sub-shot-noise reduction of fluctuations in the difference between
the photons in the single and compound fields. However, it might
happen that $ R \ge 1 $ for $ B_0 \neq B_{12} $ and $ K_{0,12} < 0
$: in this case the noise reduction factor cannot be used to check
the nonclassical nature of the measured state. However, when this
condition occurs other quantities, such as the conditional Fano
factor $ F_{c,12} $, can provide evidence of the nonclassical
character of the fields.

\section{Conclusions}

The quantum properties of two interlinked nonlinear interactions
generating a tripartite entangled state have been analyzed with
special attention to the mutual correlations in the number of
photons. In particular, we have taken into account the correlation
between the photons in the first field and the sum of photons in
the other two. The joint photon-number distribution, its
conditional photon-number distribution as well as the joint
quasi-distribution of integrated intensities have been determined
to study the nonclassical properties of the measured fields. It
has been shown that states entangled in photon numbers can occur
in the second and third fields provided that a given number of
photons is detected in the first field. In this entangled state,
sum of photon numbers in the second and third fields equals the
given number of photons in the first field and photon number in
the second field (as well as in the third field) is not
determined. The crucial role played by the quantum detection
efficiencies in the production of the conditional state is widely
discussed.

\section*{Acknowledgments}
This work was supported by projects
KAN301370701 of Grant agency of AS CR, 1M06002 and MSM6198959213
of the Czech Ministry of Education.

\end{document}